\documentclass[prd,nofootinbib,amsfonts,notitlepage]{revtex4-1}
\usepackage{graphicx,bm,amsmath,color}
\usepackage[latin1]{inputenc}
\newcommand{\be}{\begin{equation}}
\newcommand{\ee}{\end{equation}}
\newcommand{\beq}{\begin{eqnarray}}
\newcommand{\eeq}{\end{eqnarray}}

\DeclareMathOperator{\Tr}{Tr}

\begin{document}

\title{Fine-tuning problems in quantum field theory and Lorentz invariance: \\
A scalar-fermion model with a physical momentum cutoff}

\author{J.L. Cort\'es}
\affiliation{Departamento de F\'{\i}sica Te\'orica,
Universidad de Zaragoza, Zaragoza 50009, Spain}
\author{Justo L\'opez-Sarri\'on}
\email{cortes@unizar.es, jujlopezsa@unal.edu.co}
\affiliation{\it Departamento de F\'isica, Universidad Nacional de Colombia, 111321, Bogot\'a, Colombia}

\begin{abstract}
We study the consistency of having Lorentz invariance as a low energy approximation within the quantum field theory framework. A model with a scalar and a fermion field is used to show how a Lorentz invariance violating high momentum scale, a physical cutoff rendering the quantum field theory finite, can be made compatible with a suppression of Lorentz invariance violations at low momenta. The fine tuning required to get this suppression and to have a light scalar particle in the spectrum is determined at one loop.       
\end{abstract}

\maketitle

\section{Introduction}

In a theory of quantum gravity ({\bf QG}) one has to go beyond the standard notion of spacetime as a classical fixed background where different fields are defined. A new structure, quantum spacetime, should replace it and deviations from the symmetries associated to a classical spacetime (Lorentz invariance, {\bf LI}) are naturally expected. In fact arguments leading to a violation of LI are found~\cite{Ellis:1999sf, Ellis:1999jf, Gambini:1998it} in different approaches towards QG. There is also a recent proposal for a quantum field theory ({\bf QFT}) formulation of gravity based on the introduction of higher spatial derivative terms in the gravitational action~\cite{Horava:2009uw} where LI can only emerge as an approximate symmetry in the low energy limit. 
        
Although we still do not know the final form that QG will take and how LI will be realized in this theory, there is a lot of activity exploring possible observations of small deviations from LI (see for instance~\cite{Jacobson:2005bg}) and on the search for a theoretical consistent framework including them in the limit where the gravitational interaction can be neglected. The natural candidate for such a framework is an effective field theory including Lorentz invariance violating ({\bf LIV}) terms~\cite{Colladay:1998fq} but its phenomenological consistency is based on an adhoc introduction of a sufficiently strong suppression of deviations from relativistic quantum field theory ({\bf RQFT}). One has to consider either only higher (than four) dimensional LIV terms (whose effects are naturally suppressed at low energies) or coefficients of LIV terms of dimension less or equal than four proportional to inverse powers of the scale associated to LIV.


Before entering into a discussion of the consistency of an effective field theory including LIV it is necessary to point out that it is an open question whether the incorporation of departures from special relativity requires to go beyond the effective field theory framework. In fact noncommutative field theory, as an example of a QFT incorporating LIV, shows an infrared-ultraviolet mixing~\cite{Minwalla:1999px, Matusis:2000jf} violating the standard decoupling of RQFT and the problem of an insufficient suppresion of LIV at low energies is even worst~\cite{Anisimov:2001zc} than within the effective field theory framework. Another attempt to go beyond special relativity based on a generalized relativity principle~\cite{AmelinoCamelia:2000mn, Magueijo:2001cr, Magueijo:2002am, AmelinoCamelia:2002gv} involves loss of the notion of absolute locality~\cite{AmelinoCamelia:2011bm} and then a QFT implementation (if it exists) would require to go beyond the effective field theory framework also in this case.   

Within the framework of QFT, including radiative corrections due to the interactions of the standard model, it has been argued~\cite{Collins:2004bp, Collins:2006bw} that a dimension four LIV term with a coefficient proportional to the coupling of the interactions but not suppressed by inverse powers of the LIV scale is a generic consequence of radiative corrections. This is clearly inconsistent with the very stringent constrains on LIV~\cite{Jacobson:2005bg}. Possible ways to escape to this conclusion have been proposed~\cite{Gambini:2011nx}  but it has been shown~\cite{Polchinski:2011za} that they are based either on a nongeneric special property in some model of LIV or on a power suppression of LIV already present at high energies.      


On the other hand one can consider LI as an infrared atractive fixed point of a large class of LIV theories~\cite{Chadha:1982qq} but the approach to a LI low energy limit is not fast enough to be phenomenologically consistent unless one introduces some adhoc and contrived new physics~\cite{Anber:2011xf} between the energy scale associated to LIV and the energies we have explored. 
Recently the possibility to have a suppression of LIV at low energies due to a separation between the scale of validity of the effective field theory and that one of LIV has been explored~\cite{Belenchia:2016lfc} in detail in a simple toy model of a scalar field coupled to a fermion field.

All previous considerations on low energy signals of a LIV at high energies are based on the ultraviolet behaviour of QFT, which can be modified through the introduction of higher spatial derivative terms. This modification leads to an extension of the set of renormalizable interactions~\cite{Anselmi:2007ri, Anselmi:2008ry} while keeping unitarity but the insufficient suppression of LIV at low energies induced by radiative corrections seems to be also present in this case~\cite{Iengo:2009ix}. On the other hand it has been argued~\cite{Visser:2009fg} that if one follows this idea to the end by considering Lorentz symmetry breaking as a full regularization of QFT rendering it finite (``physical'' regulator) then one can safely use the tree-level approximation with a power suppression of LIV effects at low energies.     
 

In this work we will consider a QFT with an scalar and a fermionic field and the conventional interactions of RQFT but with a free Lagrangian modified by the introduction of a physical momentum scale corresponding to a generalization of Lifshitz-type QFT~\cite{Alexandre:2011kr}. This particular example will allow us to discuss in detail the validity of the different arguments used in previous discussions about the low energy signals of a LIV at high energies in a QFT. In section II we will introduce the model with a physical momentum scale. In section III we will determine the fine-tunings required to have a scalar particle and a power suppression of Lorentz invariance violations at low energies.
Finally in section IV we will present an outlook including a discussion of the relevance of the results obtained in the particular model considered in this work and some possible future directions to follow. 

\section{Scalar-fermion model}


We will consider a model with a scalar ($\phi$) and a fermionic ($\psi$) field. It is the model with the minimum number of ingredients required to have a finite theory with LIV and a simple low momentum limit.

The Lagrangian density is
\be
\tilde {\cal L} \,=\, \tilde{\cal L}_{0 \,s} + \tilde{\cal L}_{0 \,f} + \tilde{\cal L}_{int}
\label{Lagrangian}
\ee
where 
\be
\tilde{\cal L}_{int} \,=\, - \tilde y \, \bar\psi \psi \phi - \frac{\tilde\lambda}{4!} \phi^4
\label{lint}
\ee
is the standard relativistic Lagrangian density with a Yukawa ($\tilde y$) coupling and a scalar self-coupling ($\tilde\lambda$).
The free scalar and fermion Lagrangian densities are
\beq
\tilde{\cal L}_{0 \,s}  &=&  \frac{1}{2} \left(\partial_t\phi\right)^2 - \frac{1}{2} \tilde{c}_s^2  \sum_j \left(\tilde{\partial}_j \phi\right)^2 + \frac{1}{4} \tilde{m}_s^2 \tilde{c}_s^4 \phi^2 \nonumber \\
\tilde{\cal L}_{0 \,f}  &=&  i \bar\psi \gamma^0 \partial_t \psi
- i \tilde{c}_f \sum_j \bar\psi \gamma^j \tilde{\partial}_j \psi 
\label{ltilde}
\eeq
with $\partial_t$ the time derivative, ($\gamma^0$, $\gamma^j$) the Dirac matrices. Since we want a model incorporating a LIV~\footnote{For simplicity we do not include a violation of rotational invariance}, we introduce together with the mass parameter ($\tilde{m}_s$), coefficients ($\tilde{c}_s$, $\tilde{c}_f$) for the terms containing spatial derivatives $\partial_j$ of the fields
\be
\tilde{\partial}_j \phi \,=\, K_s(- \vec \partial^2/\Lambda^2) \partial_j \phi \qquad \qquad
\tilde{\partial}_j \psi \,=\, K_f(- \vec \partial^2/\Lambda^2) \partial_j \psi
\ee
with a momentum scale $\Lambda$ and real functions $K_s$, $K_f$. We will consider analytic functions in an open interval around $x=0$ and the restriction
\be
\lim_{x\to 0} K_s(x) \,=\, \lim_{x\to 0} K_f(x) \,=\, 1 \,.
\ee
so that the free theory with $\tilde{c}_s = \tilde{c}_f$ approaches in the low momentum limit $\vec p^2 \ll \Lambda^2$ a theory with a spin zero and a spin one-half particle with the same maximum velocity of propagation. This is the characteristic feature of Lorentz invariance in the absence of electromagnetic interaction which is not included in the model. We will also consider a restriction 
\be
\lim_{x\to \infty} x \, K_s^{-1}(x) \,=\, \lim_{x\to \infty} x \, K_f^{-1}(x) \,=\, 0 \,.
\ee 
In this case a generalization~\cite{Anselmi:2007ri, Visser:2009fg} of the standard power counting arguments of RQFT allows to show that the theory does not have ultraviolet divergences, therefore providing an example of a LIV physical regulator of QFT. 

A modification of the RQFT Lagrangian affecting only to terms involving spatial derivatives of the fields with no modification on the terms with time derivatives (which requires LIV) is free of the problems with unitarity and stability that appear when one considers a modification of the Lagrangian with higher time derivatives of the fields.

The model defined by the Lagrangian (\ref{Lagrangian}) has a discrete symmetry 
\be
\psi  \;  \to \; \psi^{'} \,=\, i \gamma_5 \psi {\hskip 1cm} 
\phi \; \to \; \phi^{'} \,=\, - \phi
\ee
which forbids a fermionic mass term. The sign of the coefficient of the scalar mass term has been chosen so that the discrete symmetry is spontaneously broken. Introducing a new scalar field $\varphi$ through
\begin{equation}
\phi \,=\, \tilde v + \varphi {\hskip 1cm}  {\hskip 1cm} \tilde v\,=\, \sqrt{\frac{3 \tilde{m}_s^2 \tilde{c}_s^4}{\tilde \lambda}}
\end{equation}
one can rewrite the Lagrangian (\ref{Lagrangian}) as a sum of an interaction Lagrangian 
\be
\tilde{\cal L}_{int} \,=\, - \tilde y \, \bar\psi \psi \varphi - \frac{\tilde{m}_s \tilde{c}_s^2 \sqrt{3 \tilde \lambda}}{3!} \varphi^3 - \frac{\tilde\lambda}{4!} \varphi^4
\label{varlint}
\ee   
and free scalar and fermion Lagrangians
\beq
\tilde{\cal L}_{0 \,s}  &=&  \frac{1}{2} \left(\partial_t\varphi\right)^2 - \frac{1}{2} \tilde{c}_s^2  \sum_j \left(\tilde{\partial}_j \varphi\right)^2 - \frac{1}{2} \tilde{m}_s^2 \tilde{c}_s^4 \varphi^2 \nonumber \\
\tilde{\cal L}_{0 \,f}  &=&  i \bar\psi \gamma^0 \partial_t \psi
- i \tilde{c}_f \sum_j \bar\psi \gamma^j \tilde{\partial}_j \psi - \tilde{m}_f \tilde{c}_f^2 \bar\psi \psi 
\label{varltilde}
\eeq
with $\tilde{m}_f \tilde{c}_f^2 \,=\, \tilde y \, \tilde v$.

The free parameters of the model are a scalar mass parameter $\tilde m_s$, a scalar velocity $\tilde c_s$, a fermion velocity $\tilde c_f$, a Yukawa coupling $\tilde y$, a scalar self-coupling $\tilde\lambda$ and a momentum scale $\Lambda$. Each choice of the real functions $K_s$, $K_f$ define a different model.

A good property of the model is that it includes an ingredient of the standard model of particle physics, the generation of fermion masses through spontaneous symmetry breaking. The symmetries of the model allow to show~\footnote{A detailed discussion of this point will be presented elsewhere~\cite{FW}} that the result for any S-matrix element, when one considers all momenta $\vec p$ such that $\vec p^2 \ll \Lambda^2$ and one neglects corrections proportional to positive powers of the ratio $(\vec p^2/\Lambda^2)$, can be reproduced with a perturbatively renormalizable QFT with a Lagrangian similar to (\ref{Lagrangian})-(\ref{varlint})-(\ref{varltilde}) but with $K_s = K_f = 1$. 
The low momentum effective theory is an appropriate tool to study the properties of the model in the low momentum limit including the signals of the LIV of the model in such a limit. The finiteness of the model does not preclude the appearance of ultraviolet divergences in the low momentum effective theory which are reabsorbed in their (renormalized) parameters. The parameters of the low momentum effective theory are a renormalized scalar mass parameter $m_s$, a renormalized fermion mass parameter $m_f$, a renormalized scalar velocity $c_s$, a renormalized fermion velocity $c_f$, a renormalized Yukawa coupling $y$, renormalized scalar self-couplings $\lambda_3$, $\lambda_4$ and a renormalization momentum scale $\mu$. The momentum scale $\Lambda$ is simultaneously the scale of validity of the effective theory and that one of LIV.

\section{Fine-tunings}

Let us introduce some notation. The free scalar propagator is given by
\be
\tilde{D}^{(0)}_s(p_0, \vec{p}) \,=\, \frac{i}{p_0^2 - \tilde{c}_s^2 \vec{p}^2 K_s(\vec{p}^2/\Lambda^2) - \tilde{m}_s^2 \tilde{c}_s^4}
\ee
and in the presence of interactions the full scalar propagator can be written as
\be
\tilde{D}_s(p_0, \vec{p}) \,=\, \frac{i}{p_0^2 - \tilde{c}_s^2 \vec{p}^2 K_s(\vec{p}^2/\Lambda^2) - \tilde{m}_s^2 \tilde{c}_s^4 - \tilde{\Gamma}^{(2,0)}(p_0^2, \vec{p}^2)} 
\ee
where $\tilde{\Gamma}^{(2,0)}$ is the one-particle irreducible two-scalar point function. 

The free fermion propagator is
\be
\tilde{D}_f^{(0)}(p_0, \vec{p}) \,=\, \frac{i}{\gamma^0 p_0 - \tilde{c}_f \vec{\gamma} \vec{p} \, K_f(\vec{p}^2/\Lambda^2) - \tilde{m}_f \tilde{c}_f^2}
\ee
and the full fermion propagator
\be
\tilde{D}_f(p_0, \vec{p}) \,=\, \frac{i}{\gamma^0 p_0 - \tilde{c}_f \vec{\gamma} \vec{p} \, K_f(\vec{p}^2/\Lambda^2) - \tilde{m}_f \tilde{c}_f^2 - \tilde{\Gamma}^{(0,2)}(p_0, \vec{p})} 
\ee
where $\tilde{\Gamma}^{(0,2)}$ is the one-particle irreducible two-fermion point function.

At one loop $\tilde{\Gamma}^{(2,0)}$ is a sum of three contributions, one from the tadpole diagram, one from the diagram with two scalar pro\-pa\-ga\-tors and one from the dia\-gram with two fermion propagators
\be
\begin{split}
-i \tilde{\Gamma}^{(2,0)}_1 (p_0^2, \vec{p}^2) \,=\, & -i \tilde{\lambda} \int \frac{d^4l}{(2\pi)^4} \, \tilde D_s^{(0)}(l_0, \vec{l}) + \left(-i\tilde m_s\tilde c_s^2 \sqrt{3\tilde \lambda}\right)^2 \int \frac{d^4l}{(2\pi)^4} \, \tilde D_s^{(0)}(l_0-p_0, \vec{l}-\vec{p}) \tilde D_s^{(0)}(l_0, \vec{l}) \\
& - (-i\tilde y)^2 \int \frac{d^4l}{(2\pi)^4} \Tr\left[\tilde D_f^{(0)}(l_0-p_0, \vec{l}-\vec{p}) \tilde D_f^{(0)}(l_0, \vec{l})\right]
\end{split}
\label{Gamma_1(2,0)}
\ee
In the case of $\tilde{\Gamma}^{(0,2)}$ there is only one diagram at one loop
\be
- i \tilde\Gamma^{(0,2)}_1 \,=\, (-i \tilde y)^2 \int \frac{d^4l}{(2\pi)^4} \tilde D_f^{(0)}(l_0, \vec{l}) \tilde D_s^{(0)}(l_0-p_0, \vec{l}-\vec{p})
\label{Gamma_1(0,2)}
\ee

Let us consider the scalar and fermion propagators for momenta much smaller than the momentum scale ($|\vec{p}| \ll \Lambda$). We need the first terms of the Taylor expansion for the one-particle two point functions
\be
\begin{split}
&\tilde\Gamma^{(2,0)} \,=\, \tilde\Gamma^{(2,0)}(0) + p_0^2 \frac{\partial\tilde\Gamma^{(2,0)}}{\partial p_0^2}(0) + \vec{p}^2 \frac{\partial\tilde\Gamma^{(2,0)}}{\partial\vec{p}^2}(0) + \delta\tilde\Gamma^{(2,0)} \\
&\tilde\Gamma^{(0,2)} \,=\, \frac{1}{4} \Tr\left[\tilde\Gamma^{(0,2)}(0)\right] + \frac{1}{4} \gamma^0 p_0 \Tr\left[\gamma^0 \frac{\partial\tilde\Gamma^{(0,2)}}{\partial p_0}(0)\right] - \frac{1}{4} \vec{\gamma} \vec{p} \Tr\left[\vec{\gamma} \frac{\partial\tilde\Gamma^{(0,2)}}{\partial\vec{p}}(0)\right] + \delta\tilde\Gamma^{(0,2)}
\end{split}
\ee
Neglecting corrections proportional to $(\vec{p}^2/\Lambda^2)$ one has for the inverse of the scalar propagator
\be
i \tilde D_s^{-1}(p_0, \vec{p}) \,=\, p_0^2 \left[1 - \frac{\partial\tilde\Gamma^{(2,0)}}{\partial p_0^2}(0)\right] - \vec{p}^2 \left[\tilde c_s^2 + \frac{\partial\tilde\Gamma^{(2,0)}}{\partial\vec{p}^2}(0)\right]  - \left[\tilde m_s^2 \tilde c_s^4 + \tilde\Gamma^{(2,0)}(0)\right]
\ee
At one loop the dispersion relation for the scalar particle can then be written as
\be
p_0^2 \,=\, v_s^2 \vec{p}^2 + m_s^2 v_s^4
\label{sdr}
\ee
with
\be
\begin{split}
& v_s^2 \,=\, \tilde c_s^2 + \tilde c_s^2  \frac{\partial\tilde\Gamma^{(2,0)}_1}{\partial p_0^2}(0)
 + \frac{\partial\tilde\Gamma^{(2,0)}_1}{\partial\vec{p}^2}(0) \label{vs} \\
& m_s^2 \,=\, \tilde m_ s^2 - \tilde m_s^2 \frac{\partial\tilde\Gamma^{(2,0)}_1}{\partial p_0^2} - \tilde m_s^2 \frac{2}{\tilde c_s^2}  \frac{\partial\tilde\Gamma^{(2,0)}_1}{\partial\vec{p}^2}(0) + \frac{1}{\tilde c_s^4} \tilde\Gamma^{(2,0)}_1(0)
\end{split}
\ee
The dispersion relation (\ref{sdr}) is just the dispersion relation of a relativistic particle with a maximum velocity of propagation $v_s$ and a mass $m_s$. In the expression of the squared mass of the scalar one has, together with contributions proportional to the squared mass parameter $\tilde m_s^2$, a contribution $\frac{1}{\tilde c_s^4} \tilde\Gamma^{(2,0)}_1(0)$. Using (\ref{Gamma_1(2,0)}) one finds
\be
\tilde\Gamma^{(2,0)}_1(0) \,=\, \frac{\Lambda^2}{\pi^2} \left[\frac{\tilde\lambda}{\tilde c_s} \int_0^\infty \frac{dx \, x}{K_s(x^2)} - \frac{\tilde y^2}{\tilde c_f} \int_0^\infty \frac{dx \, x}{K_f(x^2)}\right] 
+ {\cal O}(\tilde m_s^2, \tilde m_f^2)
\ee 
This means that the scalar mass will be proportional to the high momentum scale $\Lambda$ unless one has a cancellation of contributions proportional to $\Lambda^2$ in $\tilde\Gamma^{(2,0)}$. This is just the analogue of the hierarchy problem in the standard model. Although the Lorentz violating model is finite, the momentum scale $\Lambda$ acts as an ultraviolet cutoff. In order to have a scalar particle at low energies one requires a cancellation of contributions proportional to $\Lambda^2$ in the scalar propagator. This happens at one loop if
\be
\frac{\tilde\lambda}{\tilde c_s} \int_0^\infty \frac{dx \, x}{K_s(x^2)} \,=\, \frac{\tilde y^2}{\tilde c_f} \int_0^\infty \frac{dx \, x}{K_f(x^2)} \,.
\label{smcc}
\ee     
The cancellation of contributions proportional to $\Lambda^2$ is similar to the cancellation that appears in a supersymmetric theory. In both cases one has contributions with opposite signs due to bosonic and fermionic loops. In the case of a supersymmetric theory the exact cancellation is due to a symmetry under transformations among bosonic and fermionic fields leading to a relation between the Yukawa coupling that appears in the fermionic loop contribution and the scalar self-coupling that appears in the bosonic loop contribution. In the present model we also have contributions of opposite signs but the cancellation is due to a choice of the parameters of the model which should be modified order by order in the perturbative expansion.

Together with the mass of the scalar particle, another property of the model which requires a fine-tuning is the power suppression of LIV in the low momentum limit. The implementation of this fine-tuning requires the expression for the inverse of the fermion propagator when $(\vec{p}^2/\Lambda^2)\ll 1$
\be
\begin{split}
& i \tilde D_f^{-1}(p_0, \vec{p}) \,=\, \gamma^0 p_0 \left(1 - \frac{1}{4} \Tr\left[\gamma^0 \frac{\partial\tilde\Gamma^{(0,2)}}{\partial p_0}(0)\right]\right)  \\ & \qquad 
+ \vec{\gamma} \vec{p} \left(\tilde c_f + \frac{1}{4} \Tr\left[\vec{\gamma} \frac{\partial\tilde\Gamma^{(0,2)}}{\partial \vec{p}}(0)\right]\right) - \left(\tilde m_f \tilde c_f^2 + \frac{1}{4} \Tr\left[\tilde\Gamma^{(0,2)}(0)\right]\right)
\end{split}
\ee
which is proportional to 
\be
\gamma^0 p_0 + v_f \vec{\gamma} \vec{p} - m_f v_f^2 
\ee
where at one loop one has
\be
\begin{split}
& v_f^2 \,=\, \tilde c_f^2 + \frac{\tilde c_f^2}{2} \Tr\left[\gamma^0 \frac{\partial\tilde\Gamma^{(0,2)}_1}{\partial p_0}(0)\right] + \frac{\tilde c_f}{2} \Tr\left[\vec{\gamma} \frac{\partial\tilde\Gamma^{(0,2)}_1}{\partial\vec{p}}(0)\right] 
\label{vf}
\\
& m_f \,=\, \tilde m_f \left(1 + \frac{1}{2} \Tr\left[\gamma^0 \frac{\partial\tilde\Gamma^{(0,2)}_1}{\partial p_0}(0)\right] - \frac{1}{2 \tilde c_f} \Tr\left[\vec{\gamma} \frac{\partial\tilde\Gamma^{(0,2)}_1}{\partial\vec{p}}(0)\right]\right) + \frac{1}{4 \tilde c_f^2} \Tr\left[\tilde\Gamma^{(0,2)}(0)\right]
\end{split}
\ee
The interpretation of this result is that, when $(\vec{p}^2/\Lambda^2)\ll 1$, one has a relativistic fermion with a maximum velocity of propagation $v_f$ and a mass $m_f$. If one calculates $Tr\left[\tilde\Gamma^{(0,2)}_1(0)\right]$ from (\ref{Gamma_1(0,2)}) one finds that it is proportional to $\tilde m_f$. Then the mass of the fermion is proportional to the mass parameter $\tilde m_f$ and there is no contribution proportional to $\Lambda$. This could have been anticipated from the discrete symmetry that prevents to introduce a mass term for the fermion field in the Lagrangian (\ref{ltilde}).

The low momentum limit of the model is not a relativistic theory because in general the maximum velocities of propagation of the scalar ($v_s$) and the fermion ($v_f$) will be different. If we want all the effects of LIV to be  suppressed by powers of $(|\vec{p}|/\Lambda)$ then it is necessery to have $v_s=v_f$. Using the expressions for these maximum velocities in terms of the proper vertices at one loop (\ref{vs}), (\ref{vf}) and the expressions for the proper vertices at one loop (\ref{Gamma_1(2,0)}), (\ref{Gamma_1(0,2)}) we find that a power suppression of LIV at low momenta requires at one loop that
\be
\begin{split}
\tilde c_f^2-\tilde c_s^2 = & \left(\frac{\tilde y^2}{6 \pi^2 \tilde c_f}\right) \, \int_0^\infty \frac{dx}{x} \frac{\left[1 - K_f(x^2) + 2 \left(1 - K_s(x^2)\right)\right]}{K_s(x^2) \left[K_s(x^2)+K_f(x^2)\right]^2}  \\ & - \left(\frac{\tilde y^2}{4 \pi^2 \tilde c_f}\right) \, \int_0^\infty \frac{dx}{x} \frac{1 - K_f^2(x^2)}{K_f^3(x^2)}
\end{split}
\label{licc}
\ee
When the parameters $\tilde c_f$ and $\tilde c_s$ satisfy this relation there is a cancellation, at one loop and low momenta, of the LIV of the model due to the difference $(\tilde c_f - \tilde c_s)$ and the LIV due to the differences $(1 - K_s)$, $(1 - K_f)$. Note that the power suppression of LIV is not present even with the simple choice $\tilde c_s=\tilde c_f$ and $K_s=K_f$.
       
An alternative way to arrive to the same conclusion is to consider systematically the effective theory corresponding to the low momentum limit of the model. Even if the model is finite and free of any ultraviolet divergences the study of its low momentum limit by a low energy effective theory (which can be convenient for some purposes) will require the standard regularization and renormalization procedures of QFT.
A determination of the renormalized parameters $c_s$ and $c_f$ of the effective theory in terms of the parameters of the model allows to identify the condition $c_s=c_f$ as the requirement to have a powers suppression of LIV at low momenta. One can see~\cite{FW} that this leads to the relation (\ref{licc}). One could go beyond the dominant terms in the effective theory including terms of dimension higher than four that lead to contributions suppressed by powers of $(\vec{p}^2/\Lambda^2)$ at low momenta. In particular those terms could be used to study the low momentum signals of the LIV of the model.  

\section{Outlook}

Ultraviolet divergences in RQFT can be the origin of a dependence of the low energy limit on the details of the high energy theory. The absence of such a dependence (naturalness) is usually taken as a consistency requirement. But the absence of any sign of an extension of the standard model at energy scales one order of magnitud above the Higgs mass suggests to explore the possibility to go beyond naturalness accepting a fine-tuning in the parameters of the high energy theory. A signal at low energies of a LIV in the high energy theory has also been used as an argument against such violations which on the other hand are suggested by different approaches to a theory of quantum gravity. 

In this work we have introduced a simple model to illustrate the possible role of LIV as a physical regulator which alters the ultraviolet behaviour of RQFT and that appears as an ingredient in the implementation of fine-tunings, including the suppression of the signals of LIV at low energies. Instead of considering the absence of fine-tunings as a consistency requirement, we propose to use it as a guide to look for the high energy theory.

We have not considered gauge interactions. The introduction of a momentum scale in the free Lagrangean terms of a QFT to make it finite is not compatible neither with LI nor with gauge invariance. This would add a new fine-tuning problem: the condition to have a power supression of gauge invariance violations in the low momentum limit.

We have determined the fine-tunings (\ref{smcc})-(\ref{licc}) required to have a low energy theory with a scalar particle and a power suppression of LIV at one loop in a model with a scalar and a fermion field. There are many choices for the functions $K_s$, $K_f$ defining the model that satisfy the required fine-tunings. At each order in perturbation theory the required fine-tuning changes and the choice of functions $K_s$, $K_f$ has to be modified. This is the reason why we have not considered specific examples satisfying the one loop fine-tuning requirement. The model used in this work is based on a modification of the free part of a RQFT. Perhaps the introduction of a modification also in the interaction terms could lead to a simpler way to implement the fine-tuning. 



A model with a LIV physical momentum scale incorporating the fine-tuning required to have a suppression of LIV effects (and others like gauge invariant violating effects or the suppression required to have a scalar particle) in the low momentum limit can be seen as an intermediate step to look for a more fundamental theory from which such a model can be derived.            

\section*{Acknowledgments}
We would like to acknowledge the collaboration with Diego Maz\'on in the first steps of this work and several discussions with Jos\'e Manuel Carmona.
This work is supported by FPA2015-65745-P (MINECO/FEDER), Spanish DGIID-DGA Grant No. 2015-E24/2 and FONDECYT Chile Grant 1140243.

\end{document}